\documentclass{webofc}

\usepackage[varg]{txfonts}   

\begin{document}
\title{Scenarios with low mass Higgs bosons in the heavy \\ supersymmetry}

\author{\firstname{Mikhail} \lastname{Dubinin}\inst{1}\fnsep\thanks{\email{dubinin@theory.sinp.msu.ru}} \and
        \firstname{Elena} \lastname{Petrova}\inst{2,1}\fnsep\thanks{\email{petrova@theory.sinp.msu.ru}}
}

\institute{Skobeltsyn Institute of Nuclear Physics, Lomonosov Moscow State University, 119991 Moscow, Russia
\and
           Physics Department, Lomonosov Moscow State University, 119991 Moscow, Russia
          }

\abstract{Possible realistic scenarios are investigated in the minimal supersymmetric standard model (MSSM) Higgs sector extended by dimension-six effective operators. The CP-odd Higgs boson with low mass around 30--90 GeV could be consistently introduced in the regime of large threshold corrections to the effective MSSM two-doublet Higgs potential.}   

\maketitle
\section{Introduction}
\label{intro}

The discovery of the Higgs boson at the LHC in 2012 \cite{lhc-higgs} confirmed the fundamental concept of symmetry breaking in the scalar sector. Properties of a new scalar with mass 125.09$\pm$0.21(stat)$\pm$0.11(sys) GeV \cite{jhep_atlas_cms} are consistent within the statistical uncertainties with predictions of the Standard Model (SM). However, uncertainties of the signal strength and signal strength errors in most production channels are still significant and should be reduced by future analyses of ATLAS and CMS Collaborations at the energy 13 TeV. At present time their combined results at $\sqrt{s}=$7 and 8 TeV \cite{jhep_atlas_cms} for the Higgs boson production cross sections and decay rates leave a room for meaningful contributions of physics beyond the SM. In the MSSM the Higgs sector includes five Higgs bosons. In the CP conserving limit they are two $CP$-even neutral Higgs bosons $h$ and $H$, one $CP$-odd neutral Higgs boson $A$ and two charged scalars $H^\pm$ \cite{mssm}.
The mass spectrum of these scalars which is defined at the tree-level by the parameters $m_{H^\pm}$ (charged Higgs boson mass) and $\tan \beta=v_2/v_1$ (the ratio of vev's in the Higgs isodoublets),  is strongly influenced by radiative corrections coming (in the natural MSSM scenarios) from the side of the third generation SM fermions and the scalar quarks.   

Calculations of radiative corrections are performed using several approaches, which can be conditionally divided into two groups, the diagrammatic approach \cite{diagrammatic} and the effective field theory approach, the latter uses the renormalization group (RG) improved effective potential at the $m_{top}$ scale \cite{haber_hempfling,eff_potential}. Insignificant differences in the mass spectrum of scalars \cite{reconciling} evaluated with the help of diagrammatic and effective potential methods  originate from different sets of radiative corrections implemented in the codes (that applies in particular to the corrections at the two-loop level) and different renormalization schemes which are used in diagrammatic (the on-shell scheme) and effective potential (the $\overline{\rm MS}$ scheme) approaches. Besides the RG resummation of leading-logarithmic contributions, the EFT approach can be easily extended to include the threshold behaviour. The gauge coupling constants $g_1$ and $g_2$ which are measured at the $m_{top}$ scale are RG evolved to the scale of scalar quark masses ($M_S$ scale) where the MSSM boundary conditions \cite{tree-level} for the two-doublet potential are imposed and the threshold corrections are calculated, resulting in the effective parameters $\lambda_i$, $i=$1,...7 of dimension-four two-doublet potential at the $M_S$ scale. Then $\lambda_i(M_S)$ are RG evolved to the $m_{top}$ scale where the radiatively corrected Higgs boson masses are evaluated using the effective RG improved two-Higgs doublet potential. Generalization of this scheme is possible to the case of $M_S$ at the multi-TeV scale ('heavy' supersymmetry) with two thresholds, when additional mass scale $m_A$ is between $m_{top}$ and $M_S$ \cite{lee_wagner}. In calculations below we are using symbolic expressions for the effective potential parameters and normalization conventions from \cite{own}. The EFT approach specified in \cite{own} has been modified in \cite{dim_six}. In the modification \cite{dim_six} the MSSM Higgs sector is extended by 13 dimension-six operators with the effective factors in front of them $\kappa_i$, $i$=1,...13 which are calculated symbolically in the EFT. It is shown that the threshold corrections dependent on various powers of $A_{t,b}/M_S$ and $\mu/M_S$ (where $A_{t,b}$ are the trilinear Higgs-squark couplings and $\mu$ is the superfield mass parameter in the general Higgs boson-squarks interaction Lagrangian \cite{haber_hempfling}) moderately contribute to the observables, so evaluations in the range of $A_{t,b},\mu$ of the order of several or more TeV\footnote{The analysis of perturbative unitarity violation in the MSSM scalar sector deserves a separate study.} are meaningful.        

In the next Section some examples of the structure of effective couplings in front of the dimension-four and the dimension-six Lagrangian terms are given. Numerical results in the frameworks of several parametric MSSM scenarios (the benchmark scenarios \cite{benchmark}) are discussed in Section 3 with the focus on {\it low-}$m_H$ and {\it low-}$m_A$ scenarios. The question of the MSSM vacuum stability which could be relevant for a large $A_{t,b},\mu$ range in the low-mass scenarios, is discussed in Section 4.    

\section{EFT framework}
  
\begin{figure}[h]
\centering
\sidecaption
\includegraphics[width=0.6\linewidth]{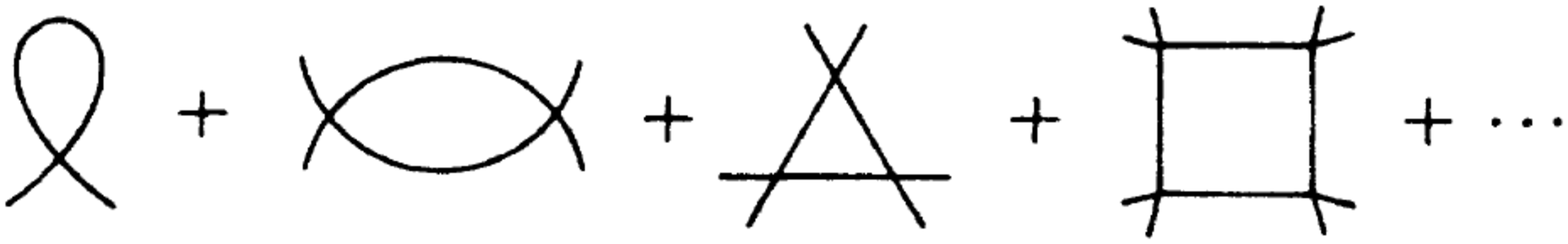}
\caption{The one-loop approximation for the effective potential in the Coleman-Weinberg framework \cite{cw73}.}
\label{fig-1}       
\end{figure}

The effective Higgs potential in the Coleman-Weinberg framework \cite{cw73} expanded to all orders of perturbation theory is shown schematically in Fig. \ref{fig-1}
\begin{equation}
\label{U}
U({\rm 1-loop})=U^{(2)}+U^{(4)}+U^{(6)}+...
\end{equation}
In the literature it is usually supposed that all higher order potential terms of effective operators which are more than dimension-four in the fields are negligibly  small if the following conditions are respected \cite{eff_potential}
\begin{equation}
\label{carenacond}
2|m_{t} A_t|<M^2_S, \qquad 2|m_{t} \mu|<M^2_S,
\end{equation}
where $m_t=$173.2 GeV is the top quark mass. Mass splitting of the squark states is small, $A_{t,b}$, $\mu$, $M_S$ parametrization of the MSSM soft supersymmetry breaking sector is most common for the MSSM benchmark scenarios.  
However, the dimension-six operators may also play a role if $A_t,\mu$ parameters satisfy the following conditions
\begin{equation}
\label{6-dim-cond}
|\mu| m_t \cot \beta \approx M_S^2, \qquad
|\mu| m_b \tan \beta \approx M_S^2, \qquad
|A_t| m_t \approx M_S^2, \qquad
|A_b| m_b \approx M_S^2, 
\end{equation}
\begin{equation}
|\mu A_t| m_t^2 \cot \beta \approx M_S^4,\qquad
|\mu A_b| m_b^2 \tan \beta \approx M_S^4, \nonumber
\end{equation}
which are true if $A_{t,b}, \, \mu$
range is of the order of a few TeV or more in combination with moderate $M_S$  at the TeV scale.
Such situation is rather unusual in the most of MSSM scenarios but estimates show that parameters $A_t,\mu$ may have so large values which are consistent with the perturbative unitarity conditions. The potential terms of the two-doublet scalar sector can be written as follows 
\begin{eqnarray}
U^{(2)} &=&
- \, \mu_1^2 (\Phi_1^\dagger\Phi_1) - \, \mu_2^2 (\Phi_2^\dagger
\Phi_2) - [ \mu_{12}^2 (\Phi_1^\dagger \Phi_2) +h.c.], \\
\label{U4}
U^{(4)} &=& \lambda_1
(\Phi_1^\dagger \Phi_1)^2
      +\lambda_2 (\Phi_2^\dagger \Phi_2)^2
+ \lambda_3 (\Phi_1^\dagger \Phi_1)(\Phi_2^\dagger \Phi_2) +
\lambda_4 (\Phi_1^\dagger \Phi_2)(\Phi_2^\dagger \Phi_1)+\\
&+& [\lambda_5/2
       (\Phi_1^\dagger \Phi_2)(\Phi_1^\dagger\Phi_2)+ \lambda_6
(\Phi^\dagger_1 \Phi_1)(\Phi^\dagger_1 \Phi_2)+\lambda_7 (\Phi^\dagger_2 \Phi_2)(\Phi^\dagger_1 \Phi_2)+h.c.], \nonumber \\
\label{U6}
U^{(6)} &=& \kappa_1 (\Phi^\dagger_1 \Phi_1)^3
+\kappa_2 (\Phi^\dagger_2 \Phi_2)^3+
\kappa_3 (\Phi^\dagger_1 \Phi_1)^2 (\Phi^\dagger_2 \Phi_2)+\kappa_4 (\Phi^\dagger_1 \Phi_1) (\Phi^\dagger_2 \Phi_2)^2+\\
&+&\kappa_5 (\Phi^\dagger_1 \Phi_1) (\Phi^\dagger_1 \Phi_2) (\Phi^\dagger_2 \Phi_1)+
\kappa_6 (\Phi^\dagger_1 \Phi_2) (\Phi^\dagger_2 \Phi_1) (\Phi^\dagger_2 \Phi_2)+ \nonumber \\
&+& [\kappa_7 (\Phi^\dagger_1 \Phi_2)^3+\kappa_8 (\Phi^\dagger_1 \Phi_1)^2 (\Phi^\dagger_1 \Phi_2)+\kappa_9 (\Phi^\dagger_1 \Phi_1) (\Phi^\dagger_1 \Phi_2)^2+ \nonumber \\
&+&\kappa_{10} (\Phi^\dagger_1 \Phi_2)^2 (\Phi^\dagger_2 \Phi_2)+
\kappa_{11} (\Phi^\dagger_1 \Phi_2)^2 (\Phi^\dagger_2 \Phi_1)+
\kappa_{12} (\Phi^\dagger_1 \Phi_2) (\Phi^\dagger_2 \Phi_2)^2+\nonumber \\
&+&\kappa_{13} (\Phi^\dagger_1 \Phi_1) (\Phi^\dagger_1 \Phi_2) (\Phi^\dagger_2 \Phi_2)+h.c.], \nonumber
\end{eqnarray}
where
\begin{equation}
\label{dublets} \Phi_i= \left(\begin{array}{c} \phi_i^+(x) \\ \phi_i^0(x) \end{array} \right)=\left(\begin{array}{c} -i \omega_i^+ \\ \frac{1}{\sqrt{2}} (v_i+\eta_i+i \chi_i) \end{array} \right), \qquad
i=1,2
\end{equation}
-- Higgs doublets with the $SU(2)$ field states and $v_1=v \cos \beta$, $v_2=v \sin \beta$ ($v=246$ GeV) -- vacuum expectation values of them. The real part of $\mu^2_{12}$ is fixed by zero eigenvalue of the mass matrix (which ensures massless Goldstone boson state and defines the $CP$-odd scalar mass $m^2_A$)
\begin{eqnarray}
\label{re12}
{\rm Re} \mu_{12}^2&=&s_\beta c_\beta [m_A^2+\frac{v^2}{2} (2 {\tt Re}\lambda_5+{\tt Re} \lambda_6 \cot\beta +{\tt Re}\lambda_7 \tan\beta)]+
v^4 \{ {\rm Re}\kappa_9 c_\beta^3 s_\beta \nonumber \\
&+&{\rm Re}\kappa_{10} c_\beta s_\beta^3
+\frac{1}{4}[{\rm Re}\kappa_8 c_\beta^4+{\rm Re} \kappa_{12} s_\beta^4+(9{\rm Re}\kappa_7+{\rm Re}\kappa_{11}+{\rm Re}\kappa_{13}) s_\beta^2 c_\beta^2] \},
\end{eqnarray}
where $s_\beta=\sin \beta$, $c_\beta=\cos \beta$ and so on. The charged Higgs boson mass is shifted from the tree-level value by $\Delta \lambda_{4,5}$ and $\kappa_i$, $i$=5,6,7,9,10,11
\begin{eqnarray}
\label{mcharged}
m_{H^\pm}^2&=&m_W^2+m_A^2-\frac{v^2}{2} ({\rm Re}\Delta \lambda_5-\Delta \lambda_4) \nonumber\\
&+& \frac{v^4}{4} [c_\beta^2(2 {\rm Re}\kappa_9-\kappa_5)+s_\beta^2(2 {\rm Re}\kappa_{10}-\kappa_6)-s_{2 \beta} ({\rm Re}\kappa_{11}-3{\rm Re}\kappa_7)] .
\end{eqnarray}
Tree-level boundary conditions for the Higgs self-couplings at the scale $M_S$ are \cite{tree-level}
\begin{eqnarray}
\label{lts}
\lambda_{1,2}^{\tt tree}({M_{S}})&=&\frac{g_1^2+g_2^2}{4}, \qquad 
\lambda_{3}^{\tt tree}({M_{S}})=\frac{g_2^2-g_1^2}{4}, \qquad
\lambda_4^{\tt tree}({M_{S}}) = \frac{g_2^2}{2},\\
&&\lambda_{5,6,7}^{\tt tree}({M_{S}})=0, \qquad \kappa_{1,...,13}^{\tt tree}({M_{S}})=0,
\end{eqnarray}
where $g_1$ and $g_2$ are the standard gauge couplings, while at the loop-level and below $M_S$ scale these couplings acquire radiative corrections

\begin{center}
$
\lambda_{1,2}(M)=\lambda_{1,2}^{\tt tree}({M_{S}})-\Delta \lambda_{1,2}(M)/2, \qquad
\lambda_{3,...,7}(M)=\lambda_{3,...,7}^{\tt tree}({M_{S}})-\Delta \lambda_{3,...,7}(M),
$\\
$
\kappa_{1,...13}(M)=\Delta \kappa_{1,...13}(M).
$
\end{center}
Radiative corrections to parameters $\lambda_i$, $i$=1,...7, in the effective field theory framework have been analysed in ref. \cite{eff_potential, haber_hempfling, own}. Radiative corrections to the parameters $\kappa_i$, $i=$1,...13 in the approximation of degenerate squark masses have been obtained in \cite{dim_six}.
An example of the one-loop RG-improved threshold correction structure for $\lambda_1$ and the threshold correction for $\kappa_1$ in the form which uses $A_{t,b,}/M_S$ and $\mu/M_S$ power terms is
\begin{eqnarray}
-\frac{\Delta \lambda_1^{\rm thr}}{2} &=& \frac{3}{32\pi^2} \Big[ h^4_b \frac{|A_b|^2}{M^2_{S}}\left(2-\frac{|A_b|^2}{6 M^{\,2}_{S}}\right)
-h^4_t\frac{|\,\mu|^4}{6 M^{\,4}_{S}} 
 + \,2 h_b^4 l +\,\frac{g_2^2+g_1^2}{4\,M^{\,2}_{S}}(h^2_t|\,\mu|^2-h^2_b|A_b|^2) \Big]  \\
&+& \, \frac{1}{768\pi^2}\,
\left(11 g_1^4 + 9g_2^4 - 36 \,(g_1^2+g_2^2)\,h_b^2\right) l , \nonumber  \\
\Delta \kappa_1^{\rm thr} &=& \frac{h_b^6}{32 M_S^2 \pi^2} \left(2-\frac{3 |A_b|^2}{M_S^2}+\frac{|A_b|^4}{M_S^4}-\frac{|A_b|^6}{10 M_S^6} \right)
- h_b^4 \frac{g_1^2+g_2^2}{128 M_S^2 \pi^2} \left( 3-3\frac{|A_b|^2}{M_S^2}+\frac{|A_b|^4}{2 M_S^4} \right) \\
&+& \frac{h_b^2}{512 M_S^2 \pi^2} \left(\frac{5}{3}g_1^4+2 g_1^2 g_2^2+3g_2^4 \right) \left(1-\frac{|A_b|^2}{2 M_S^2} \right) 
-h_t^6 \frac{|\mu|^6}{320 M_S^8 \pi^2} \nonumber\\
&+& h_t^4 \frac{(g_1^2+g_2^2)|\mu|^4}{256 M_S^6 \pi^2}
-h_t^2 \frac{(17 g_1^4-6g_1^2 g_2^2+9g_2^4) |\mu|^2}{3072 M_S^4 \pi^2}
+\frac{g_1^2}{1024 M_S^2 \pi^2} (g_1^4-g_2^4), \nonumber
\end{eqnarray}
where $l\equiv\ln\left(\frac{ M^{\,2}_{S}}{\sigma^2}\right)$, $\sigma=m_{top}$ is the renormalization scale, $h_{t}=\frac{g_2 m_{t}}{\sqrt{2}m_W \sin \beta}$ and $h_{b}=\frac{g_2 m_{b}}{\sqrt{2}m_W \cos \beta}$ are the Yukawa couplings. One can notice inspecting such explicit forms that radiative corrections $\Delta \kappa^{\rm thr}$ begin to play a role if the conditions (\ref{6-dim-cond}) are true.
In this case the corrections coming from the dimension-six operators of the one-loop potential can be appreciable even if the inequalities (\ref{carenacond}) are respected.

\section{Implications under the benchmark scenarios {\it low-}$m_H$ and {\it low-}$m_A$}
\label{sec-1}

\begin{figure}[hbtp]
\begin{minipage}[h]{0.5\linewidth}
\center{\includegraphics[width=0.8\linewidth]{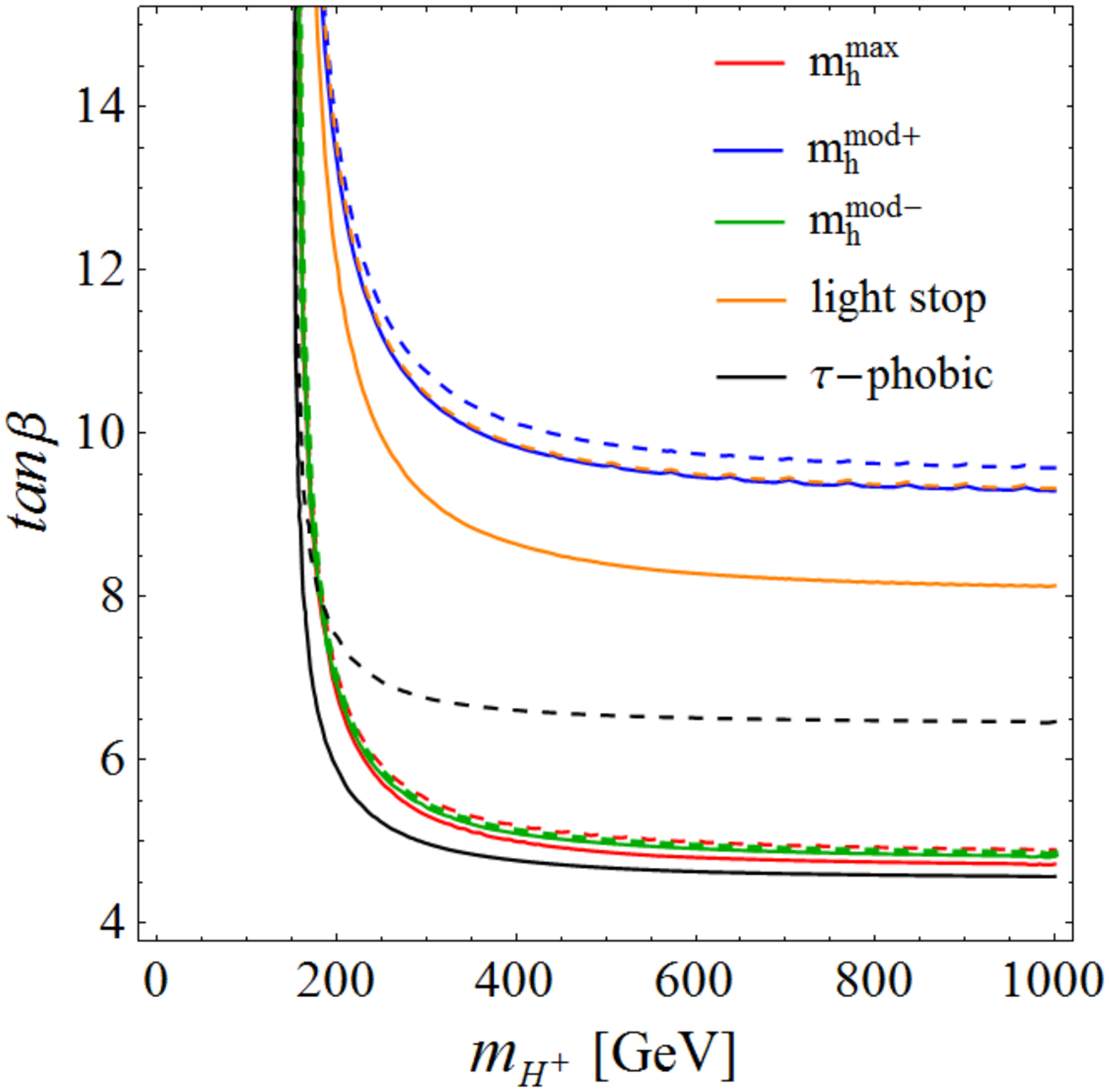}}  \\(a)
\end{minipage} 
\hfill
\begin{minipage}[h]{0.5\linewidth}
\center{\includegraphics[width=0.8\linewidth]{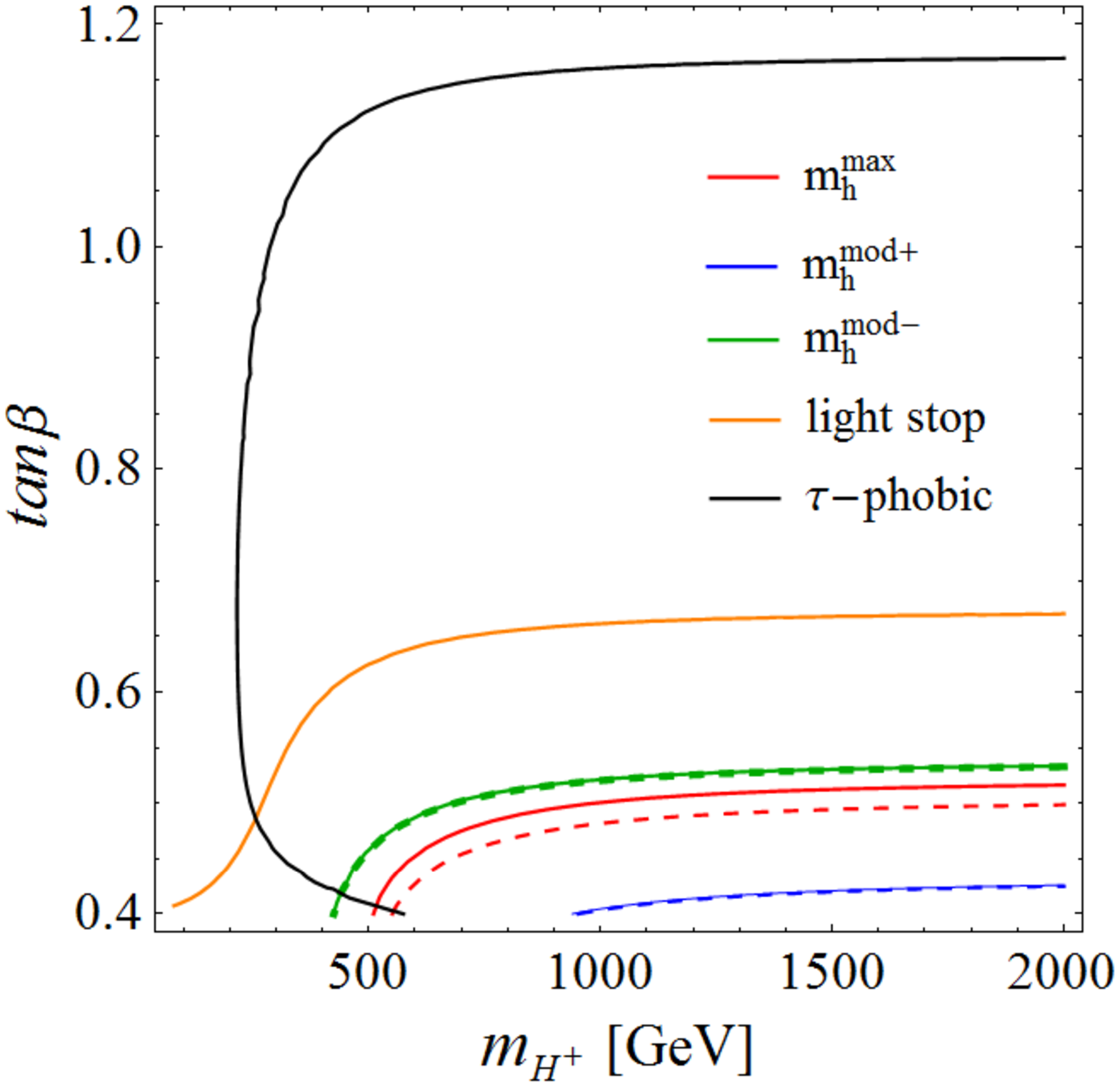}} \\(b)
\end{minipage} 
\caption{Isocontours of the lightest Higgs boson mass $m_h=125$ GeV in $(m_H^\pm, \tan \beta)$-plane for various benchmark scenarios, see Table \ref{tabl:carena}. Solid lines - Lagrangian terms of dimension-four are accounted for, dashed lines correspond to calculation which takes into account radiative corrections coming from additional dimension-six Lagrangian terms. Right panel - isocontours of $m_h=125$ GeV at a low $\tan \beta$.}
\label{ris1}
\end{figure}

\begin{figure}[h]
\centering
\sidecaption
\includegraphics[width=0.6\linewidth]{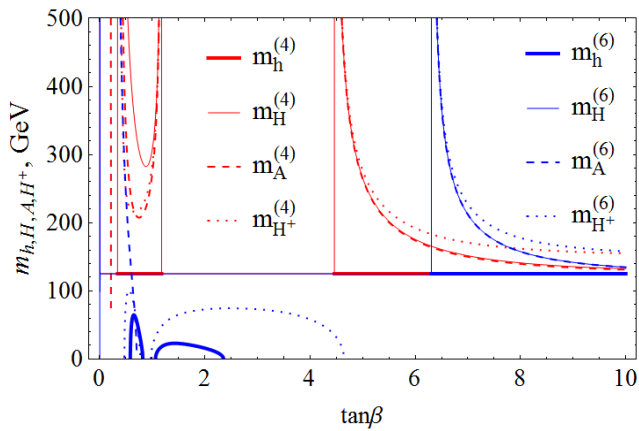}
\caption{Higgs boson masses for $\tau$-{\it phobic} scenario, see Table \ref{tabl:carena}. Red lines correspond to Higgs boson masses calculated with dimension-four Lagrangian terms including effective parameters $\lambda_i$, $i=$1,...7. Blue lines correspond to a calculation with additional dimension-six potential terms \cite{dim_six} which include effective parameters $\kappa_i$, $i=$1,...13. Due to condition $m_h$=125 GeV, all Higgs boson masses are analytically defined (see \cite{dim_six}). In the low $\tan \beta$ range, tachyonic states of Higgs bosons appear.}
\label{fig-2}       
\end{figure}

Experimental constraints on the MSSM parameter space are imposed by the search for a non-standard bosons in the channels $\tau^+ \tau^-$, $b \bar b$ and $\tau^\pm \nu_\tau$, either inclusive or with $b$-jets tagging. As mentioned above, 
calculations of observables in the MSSM parameter space are usually performed in a number of benchmark scenarios which reduce the full MSSM parameter space to a degenerate versions where the number of free parameters is five or six (if a phase of explicit CP violation is accounted for). In the natural MSSM scenarios of this sort radiative corrections depend on the superpartner mass scale $M_S$, the parameters of the soft supersymmetry breaking terms $A_{t,b}$ and the Higgs superfield mass parameter $\mu$ which are fixed to specific benchmark sets which demonstrate different consequences for the LHC phenomenology. Seven benchmark scenarios are specified in \cite{benchmark}, see the parameter sets in Table \ref{tabl:carena}. With the parameters  $M_S$, $A_t=A_b$ and $\mu$ fixed to benchmark values only $m_{H^\pm}$ and $\tan \beta$ are changed, so various constraints and exclusion contours are displayed in the ($m_{H^\pm}$, $\tan \beta$) plane. For our purposes it is also interesting to fix $m_{H^\pm}$, $\tan \beta$ and reconstruct the mass contours in the ($A_t$,$\mu$) plane.
Corrections to the Higgs boson mass spectrum which are induced by dimension-six effective operators are shown in Fig.\ref{ris1} and Fig.\ref{fig-2}. Note that in addition to medium and high $\tan \beta$ contours usually referred in the literature and used for the analyses (see e.g. \cite{CMS}), acceptable contours at $\tan \beta <$1 are observed (see also \cite{hxswg}) in the scenarios $m^{max}_h$, $m^{mod+}_h$, $m^{mod-}_h$, {\it light stop}, {\it light stau} and $\tau$-{\it phobic}. It is assumed that a rough lower bound $\tan \beta \geq m_t/600$ GeV \cite{barger90} is respected.
Corrections from dimension-six operators in the two-Higgs doublet sector substantially shift the contours in a number of scenarios reducing as a rule the acceptable regions of the MSSM parameter space.


\begin{table}[h!]
\label{tabl}
\begin{center}
\begin{tabular}{cccccccc}
\hline
 & $m_h^{\rm max}$ & $m_h^{\rm mod+}$ & $m_h^{\rm mod-}$ & light stop & light stau & $\tau$-phobic & low-$M_H$ \\
\hline
$m_{\rm top}$ [GeV] &  &  &  & 173.2 &  &  & \\
$M_{S}$ [GeV] & 1000 & 1000 & 1000 & 500 & 1000 & 1500 & 1500\\
$\mu$ [GeV] & 200 & 200 & 200 & 350 & 500 & 2000 & varied\\
$X_t^{\overline{\rm MS}}/M_{S}$ & $\sqrt{6}$ & 1.6 & -2.2 & 2.2 & 1.7 & 2.9 & 2.9\\
\hline
\end{tabular}
\caption{Parameter sets for benchmark scenarios, see \cite{benchmark}.}
\label{tabl:carena}
\end{center}
\end{table}

The case of catastrophic changes of the mass spectrum in the $\tau$-{\it phobic} scenario is shown in Fig.\ref{fig-2}. In the $\tan \beta$ range from 1 to 4.5 $h$(125 GeV) is replaced by $H$(125 GeV) and $A$-state becomes tachyonic. The state at $m_h=$125 GeV for low $\tan \beta$ in the normally ordered mass spectrum which follows from the dimension-four effective potential does not exist if the thirteen dimension-six terms are introduced. This is a typical picture in the regions of parameter space, where $\tan \beta \approx 0.4-1$ and $m_{H^+} \geq 500$ GeV. 

Besides a natural possibility to identify 125 GeV state as the lightest CP-even Higgs boson $h$, more exotic possibility  to interpret the observed state as the heavy CP-even Higgs boson $H$ of the MSSM has been analysed \cite{lowmh}. Such scenario is known as {\it low-}$m_H$. Alignment of the $H$ boson couplings to fermions and bosons (the $H$-alignment limit) when the couplings of $H$ are the SM-like respects the approximate equality of mixing angles $\alpha \sim \beta$, which is different from the $h$-alignment limit when $\alpha-\beta \sim \pi/2$\footnote{Let us remind one that $h$-alignment limit \cite{alignment} was proposed in the context of an assumption that the observed Higgs boson is the lightest $CP$-even scalar $h$.}. The alignment limit of the MSSM \cite{alignment} where couplings of $h$ are close to the SM values, is observed only for intermediate $\tan \beta$ at large masses of $m_H$, $m_A$ and $m_{H^\pm}$ in a majority of scenarios.

It was shown \cite{benchmark} that in the case of dimension-four effective Higgs potential terms in the fields $U^{(4)}$, Eq.(\ref{U}) and for {\it low-}$m_H$ benchmark scenario (see Table \ref{tabl:carena}) with fixed value of $CP$-odd scalar mass $m_A =$110 GeV
the lightest Higgs boson mass is changing in an interval $77$ GeV $\leq m_h \leq 102$ GeV. The same behaviour is observed in our case with results for masses insignificantly higher. If $U^{(6)}$ terms are added, radiative corrections from dimension-six effective operators \cite{dim_six} decrease the values of masses by 2--3 GeV. The decoupling limit is poorly realized in {\it low-}$m_H$ where all Higgs boson masses are about 100 GeV. So with $m_A =$110 GeV the $H$-alignment limit $\alpha \sim \beta$ seems to be not realistic and in the following we shift the CP-odd boson mass $m_A$ to lower values of 30 GeV and 90 GeV identifying 125 GeV state as either the CP-even scalar $h$ or the CP-even scalar $H$ ({\it low-}$m_A$ scenario). 

\begin{figure*}
\centering
\begin{minipage}[h]{0.44\linewidth}
\center{\includegraphics[width=0.9\linewidth]{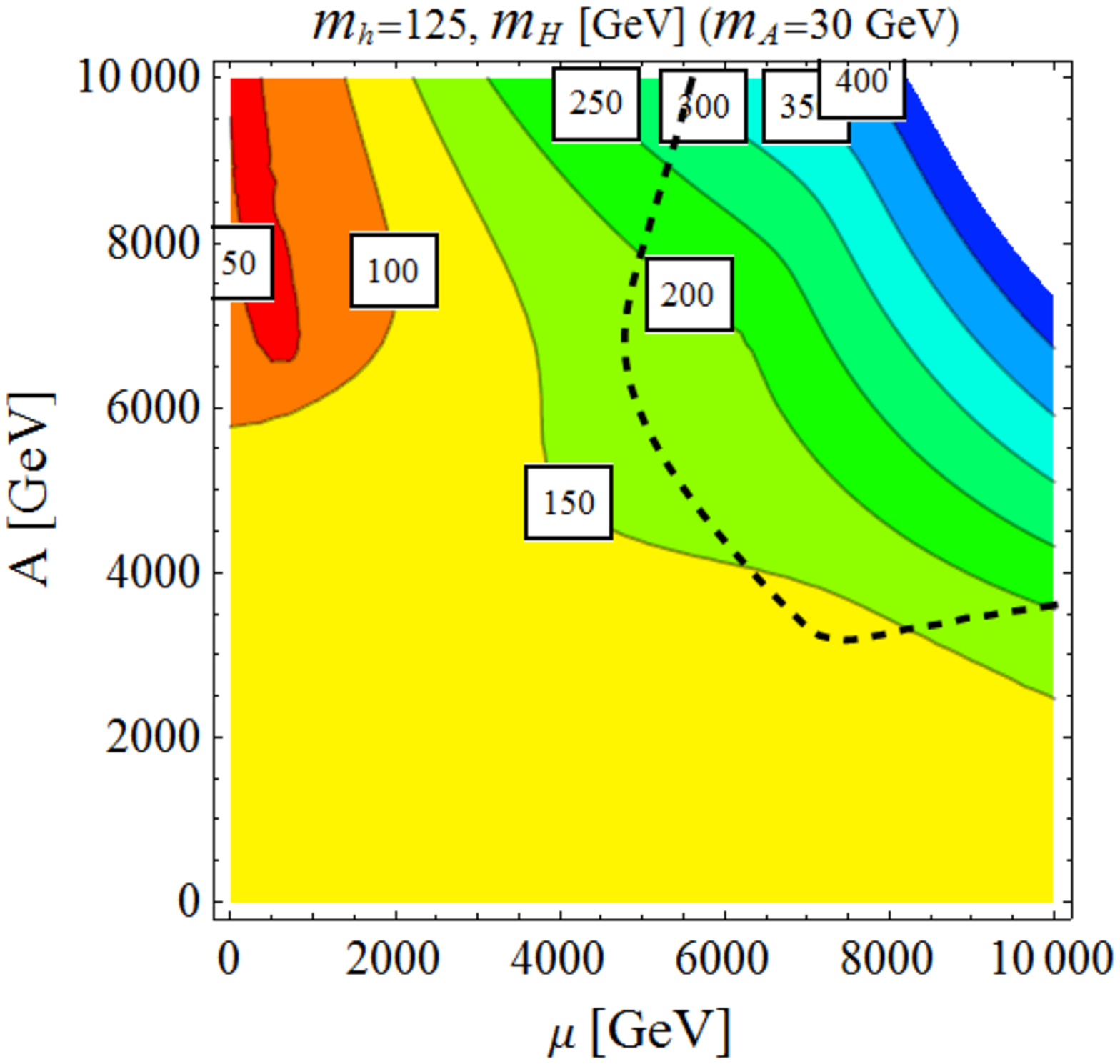}}\\(a)
\end{minipage}
\begin{minipage}[h]{0.44\linewidth}
\center{\includegraphics[width=0.94\linewidth]{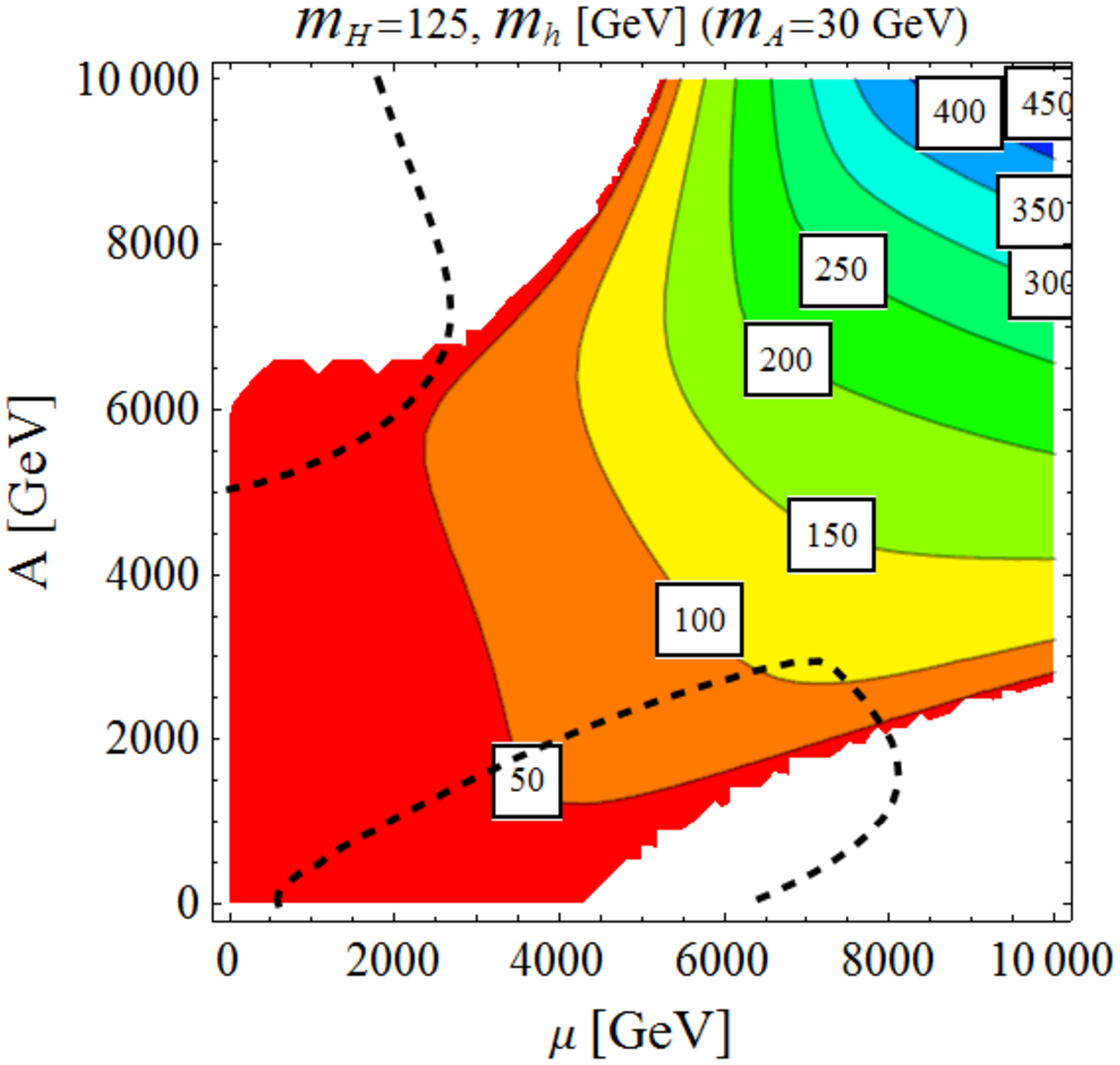}}\\(b)
\end{minipage}
\begin{minipage}[h]{0.44\linewidth}
\center{\includegraphics[width=0.9\linewidth]{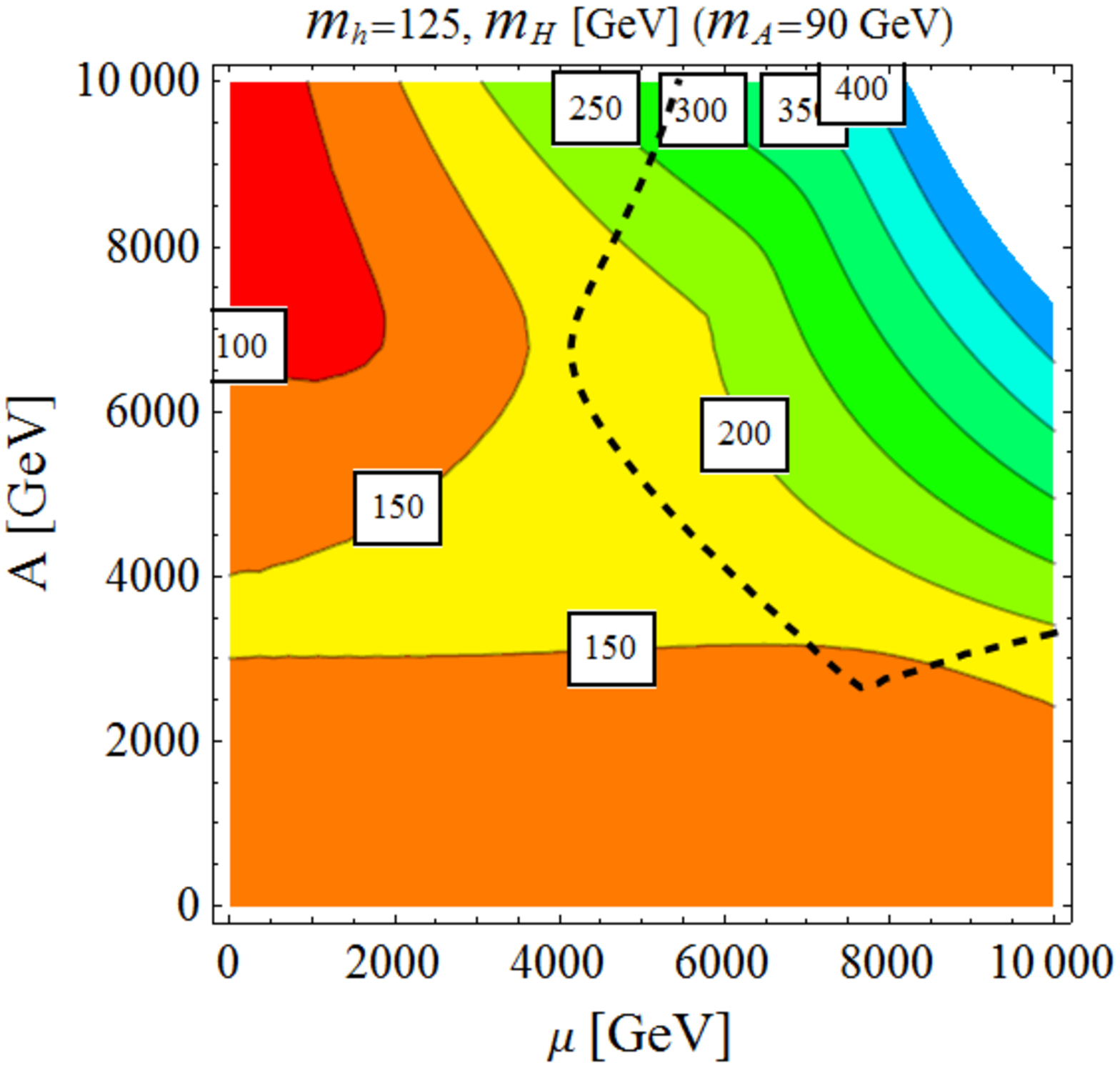}}\\(c)
\end{minipage}
\begin{minipage}[h]{0.44\linewidth}
\center{\includegraphics[width=0.94\linewidth]{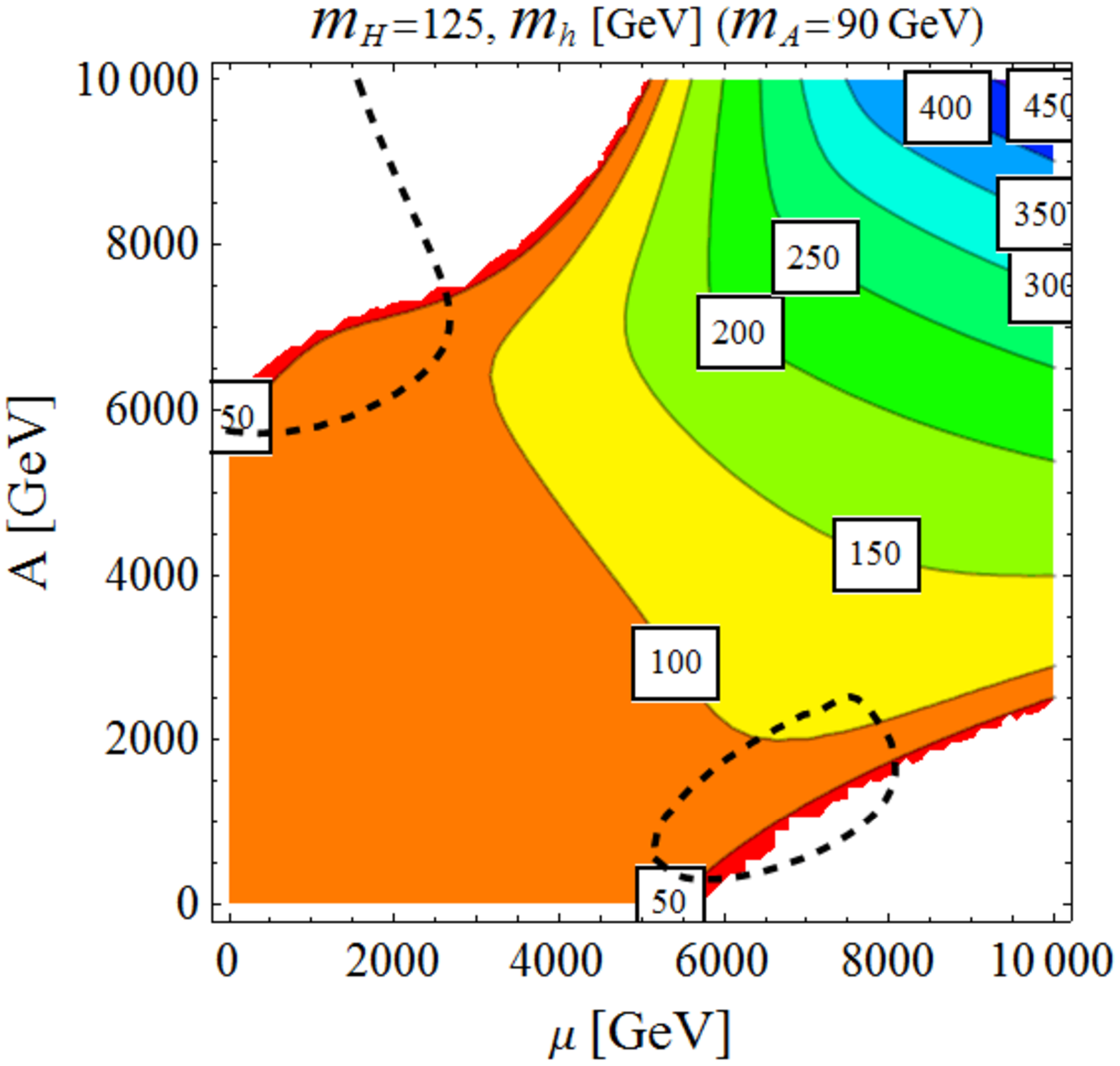}}\\(d)
\end{minipage}
\begin{minipage}[h]{0.44\linewidth}
\center{\includegraphics[width=0.9\linewidth]{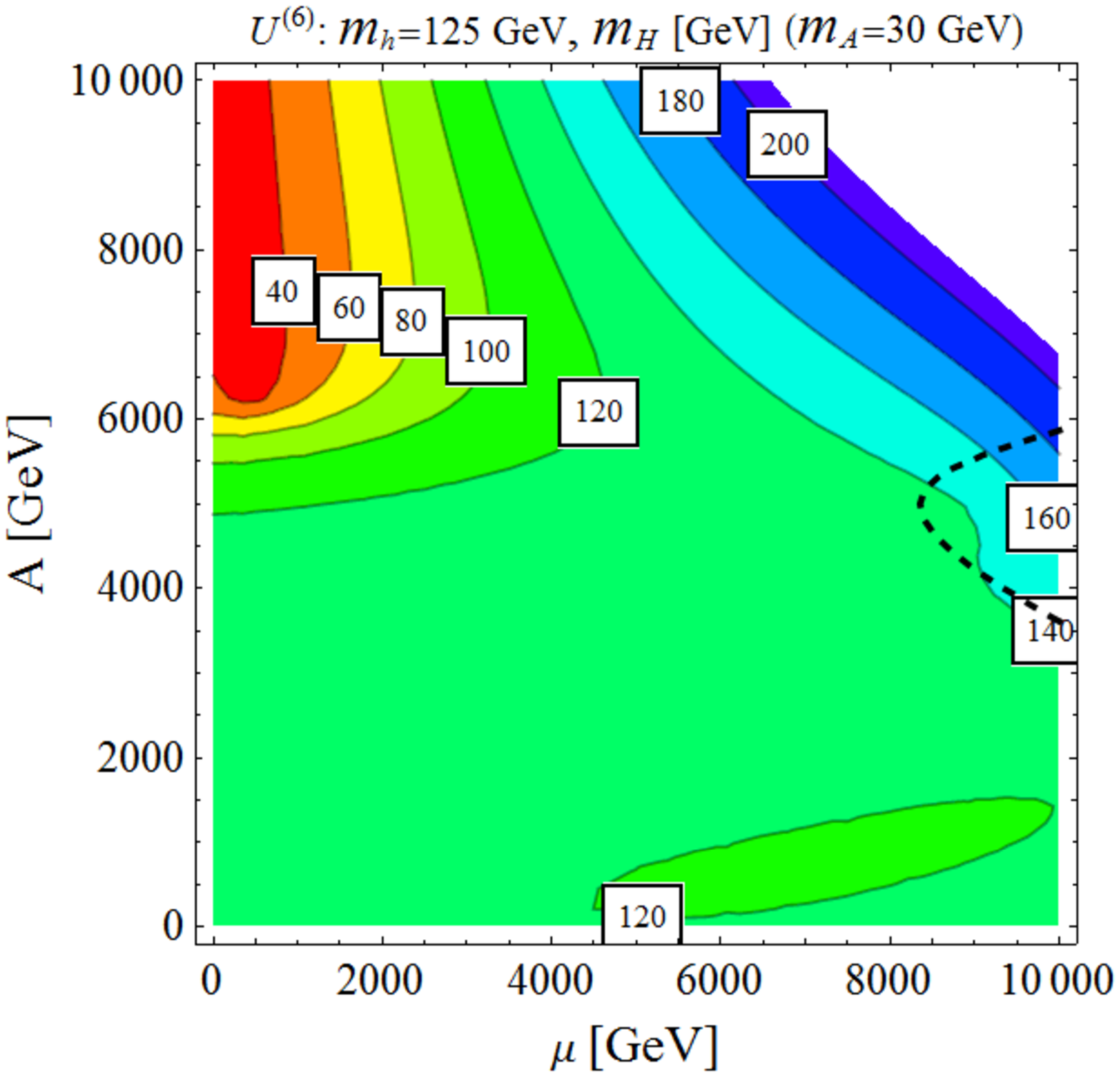}}\\(e)
\end{minipage}
\begin{minipage}[h]{0.44\linewidth}
\center{\includegraphics[width=0.94\linewidth]{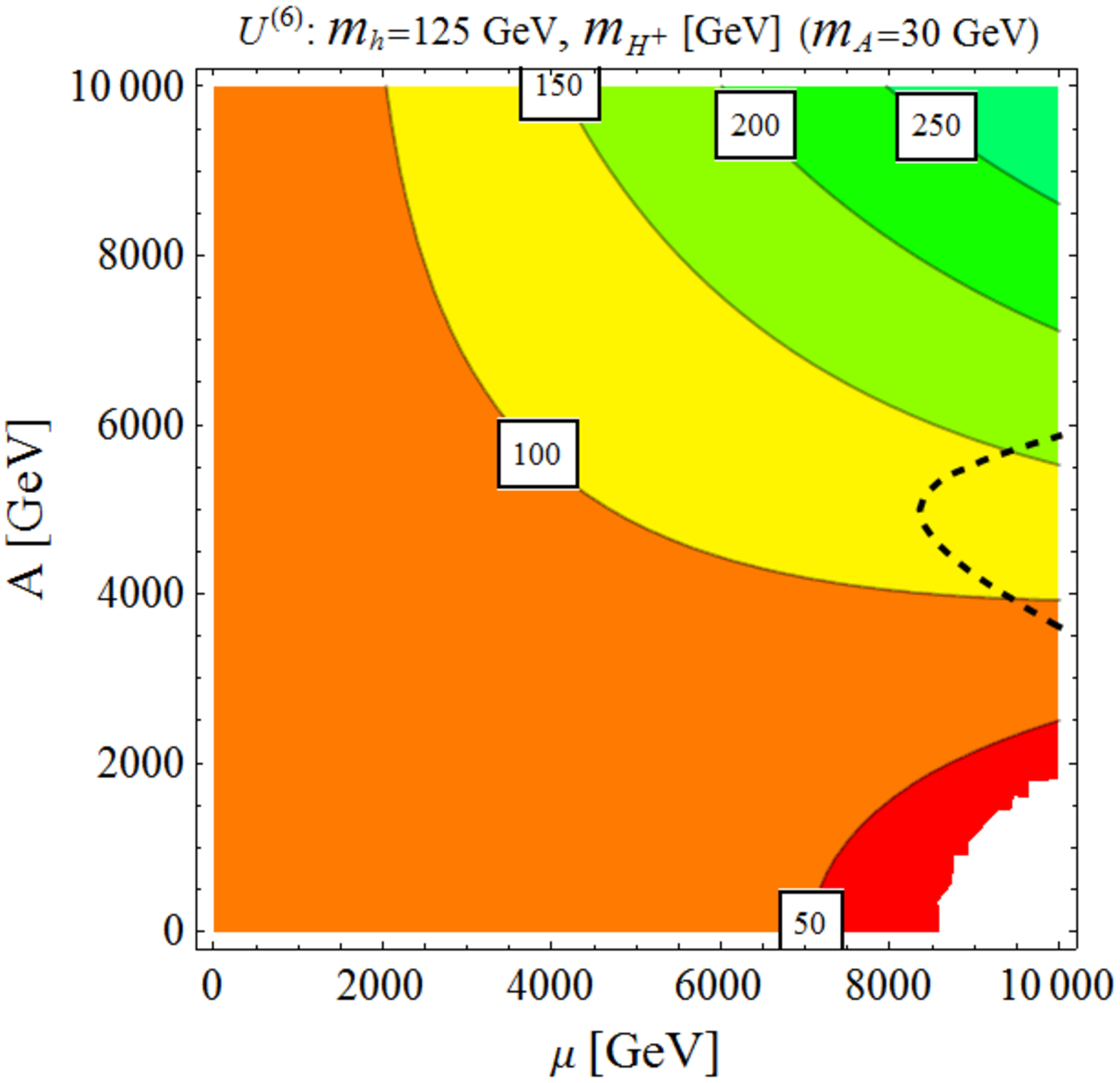}}\\(f)
\end{minipage}
\vskip -5mm
*\vspace*{-2mm}       
\caption{Isocontours for the Higgs boson masses reconstructed for {\it low-}$m_H$ and {\it low-}$m_A$ scenarios in the ($\mu$, $A_{t,b}$) plane: 
(a) $m_A$=30 GeV is fixed, various $m_H$ contours are superimposed on the $m_h=$125 GeV contour marked by dashed line;
(b)  $m_A$=30 GeV is fixed, various $m_h$ contours are superimposed on the $m_H=$125 GeV contour marked by dashed line;
(c)  the same configuration as (a), but $m_A$=90 GeV;
(d)  the same configuration as (b), but $m_A$=90 GeV. Everywhere $M_S$=1500 GeV, $\tan \beta$=2. In the cases (a) and (c) the alignment limit is respected, the charged Higgs boson mass is 150-170 GeV. In the cases (b) and (d), which are much less reasonable, the alignment limit does not take place in the full parameter space and the charged Higgs boson mass is 90--100 GeV. Plot (d) and plot (f) show the isocontours of $m_H$ and $m_{H^\pm}$ at $m_A=$30 GeV, $M_S=$1500 GeV and $\tan \beta=$5 when $m_h=$125 GeV (solid dashed line).}
\label{fig-4}       
\end{figure*}

\begin{figure*}
\centering
\begin{minipage}[h]{0.44\linewidth}
\center{\includegraphics[width=0.9\linewidth]{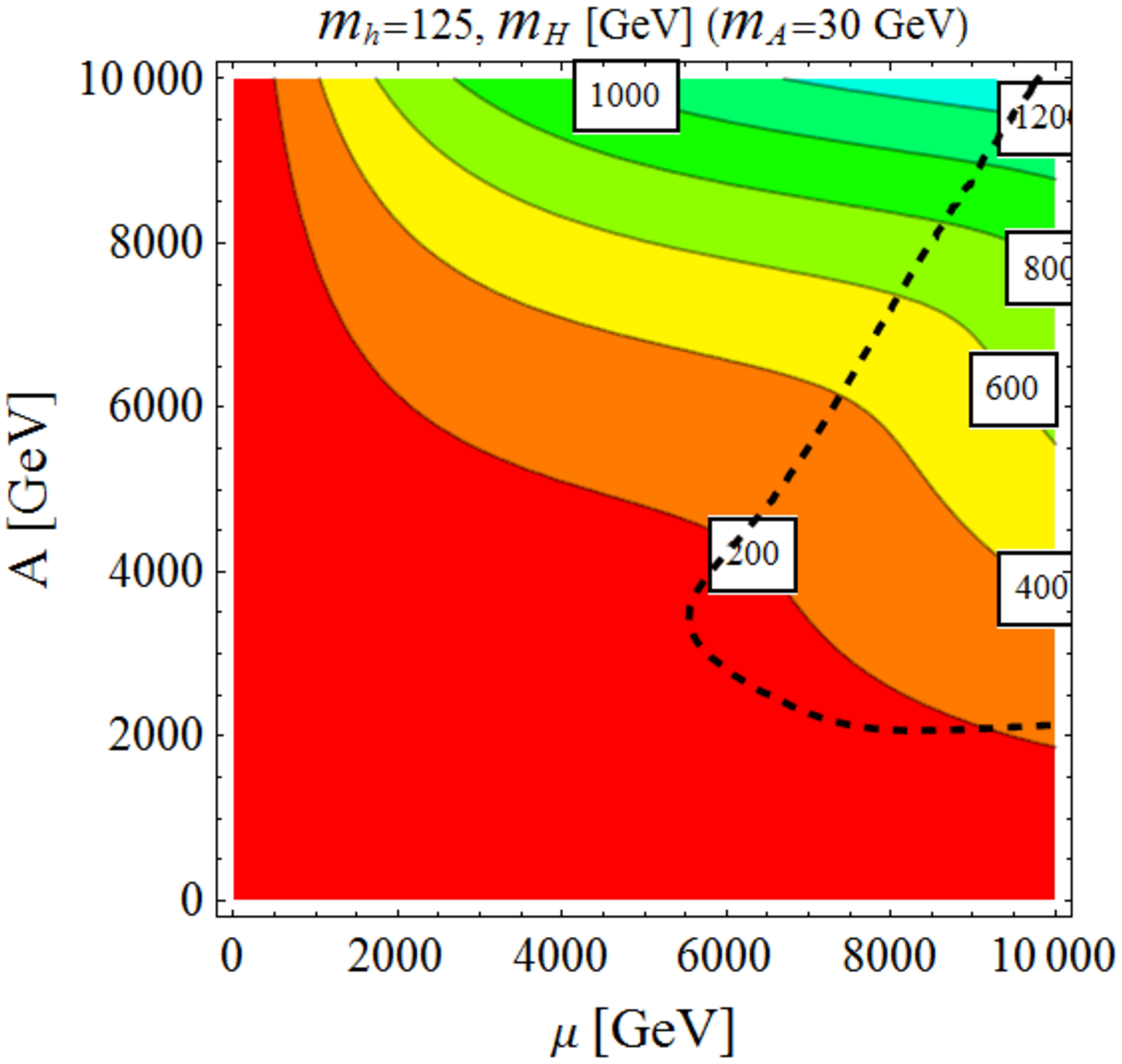}}\\(a)
\end{minipage}
\begin{minipage}[h]{0.44\linewidth}
\center{\includegraphics[width=0.94\linewidth]{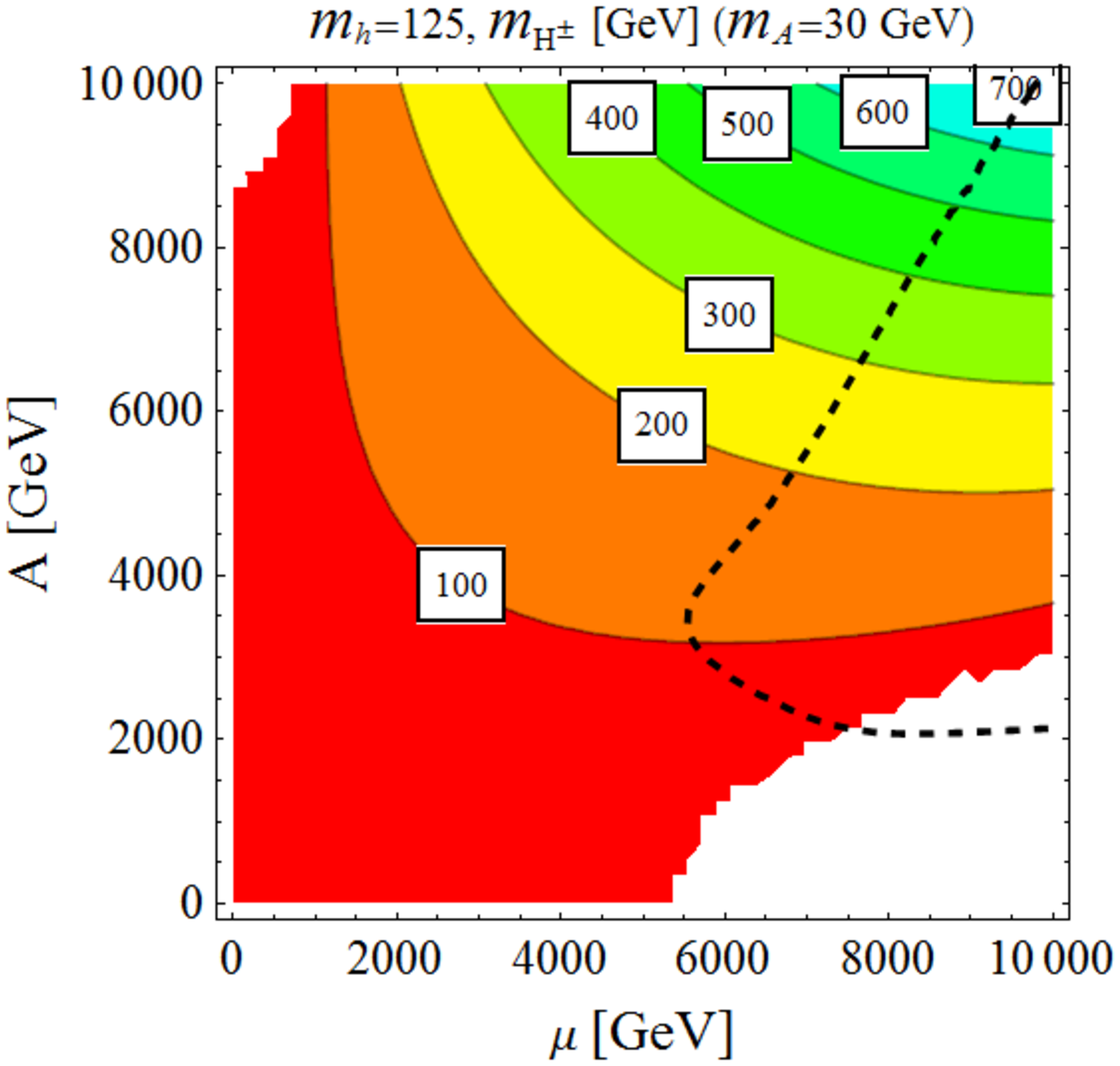}}\\(b)
\end{minipage}
\begin{minipage}[h]{0.44\linewidth}
\center{\includegraphics[width=0.9\linewidth]{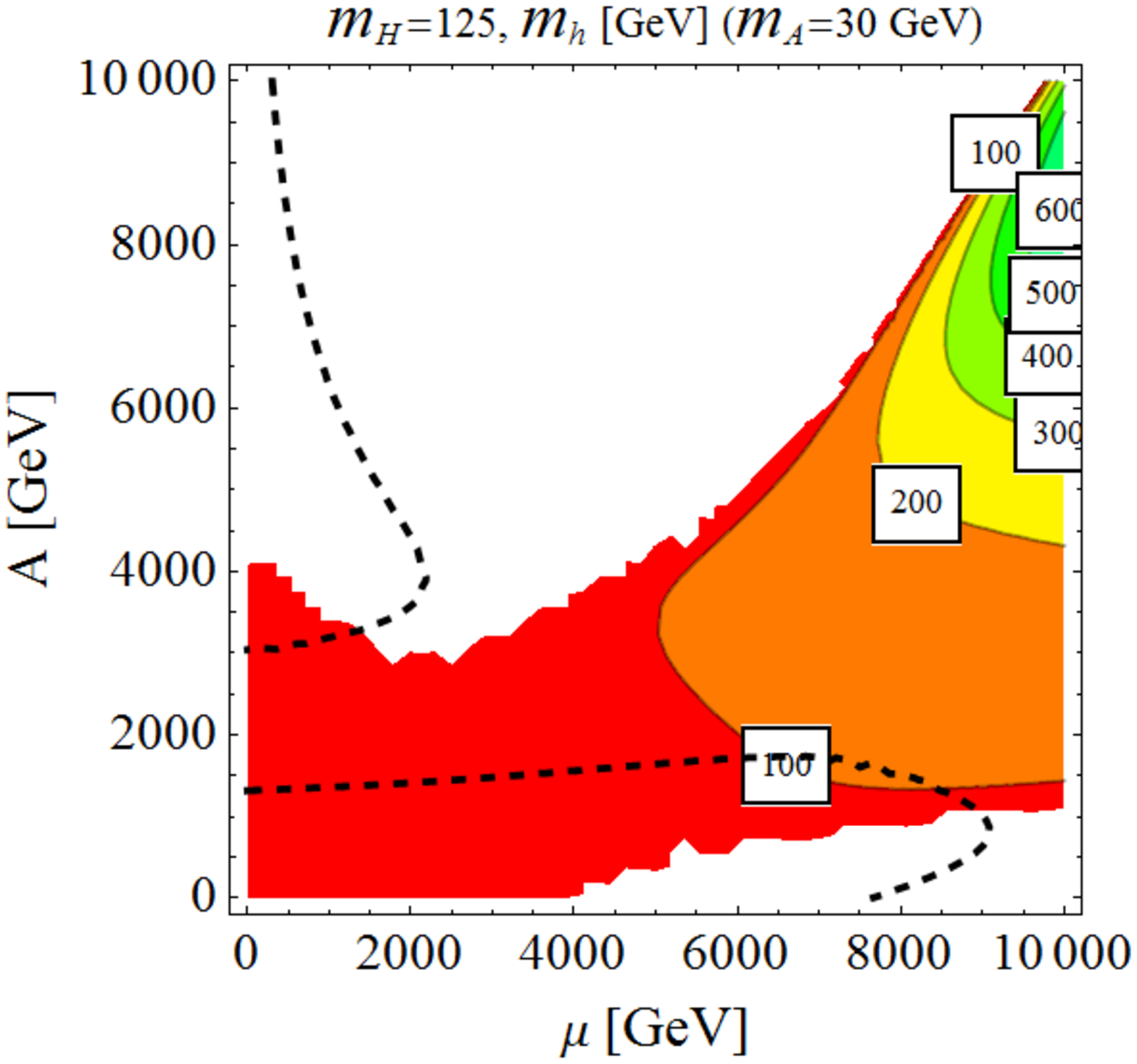}}\\(c)
\end{minipage}
\begin{minipage}[h]{0.44\linewidth}
\center{\includegraphics[width=0.94\linewidth]{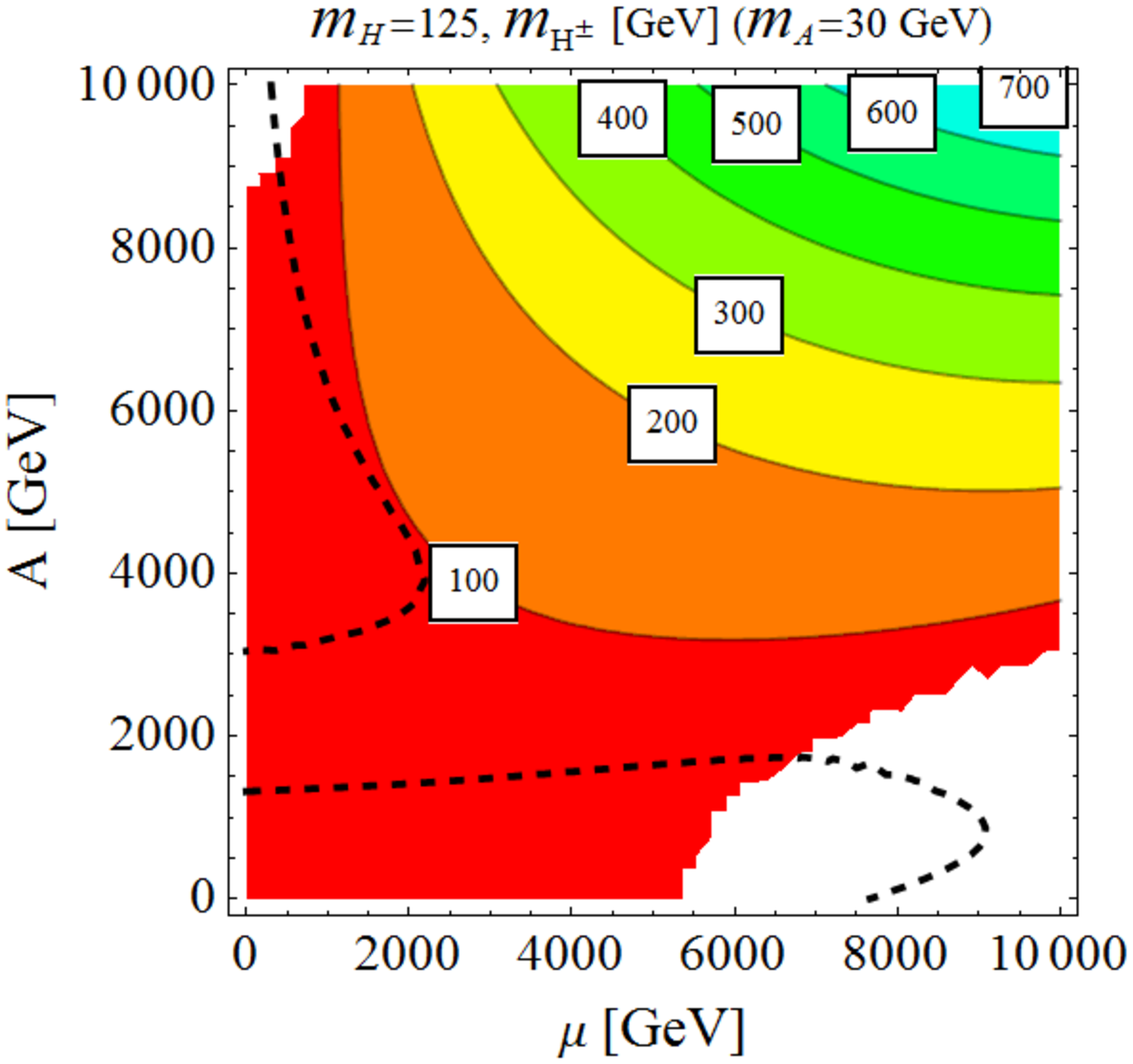}}\\(d)
\end{minipage}
\vskip -5mm
*\vspace*{-2mm}       
\caption{Isocontours for the Higgs boson masses reconstructed for {\it low-}$m_H$ and {\it low-}$m_A$ scenarios in the ($\mu$, $A_{t,b}$) plane:
$m_A$=30 GeV, $M_S=$1000 GeV and $\tan \beta=$5,
the contour marked by dashed line corresponds to 125 GeV, 
(a) various $m_H$ contours are superimposed on the $m_h=$125 GeV contour marked by dashed line;
(b) various $m_{H^\pm}$ contours are superimposed on the $m_h=$125 GeV contour marked by dashed line;
(c, d)  the same configuration as (a, b), but $m_H$=125 GeV.
}
\label{fig-4a}       
\end{figure*}

Experimental signs of the low mass $A$-boson state are not discovered, although the search is in progress \cite{cms_peak}. Typical situation at the large values of $A_{t,b}$ and $\mu$ parameters which are approaching 5 TeV in the scenario with low $m_A$ value of 30 GeV is shown in Fig.\ref{fig-4}(a), where along the dashed solid black line $m_h=$125 GeV.  Isocontours of $m_H$ demonstrate possible mass range of 300--400 GeV. Charged Higgs boson is sufficiently heavy with mass up to 170 GeV. The $h$-alignment limit takes place in this case at $\tan \beta$ from 2 to 5 and $M_S$ of around 2 TeV. In Fig.\ref{fig-4}(b) we swap $h$ and $H$, so contours of constant $m_h$ for the case when $m_H=$125 GeV (dashed solid black line) are consistent for masses less than 100 GeV. In this case the charged Higgs boson mass is in the interval 90--100 GeV and $H$-alignment limit is not respected. Analogous plots for the case $m_A=$90 GeV are shown in Fig.\ref{fig-4}(c), where $h$-alignment limit is respected, $m_{H^\pm}$ is around 150-170 GeV, and Fig.\ref{fig-4}(d), where $H$-alignment limit is not respected, $m_{H^\pm}$ is around 100 GeV.
The mass $m_{H^\pm}$ can achieve values of around 700 GeV if $m_h$=125 GeV, $M_S<$ 1500 GeV, $\tan \beta$ of around 3--6, $A_t$ and $\mu$ of around 10 TeV, see Fig.\ref{fig-4a}.
In Fig.\ref{fig-4}(e),(f) one can see the isocontours for $m_H$ (left panel) and charged Higgs mass $m_{H^\pm}$ (right panel) in {\it low-}$m_A$ scenario, where the $h$-alignment limit is available. 

\section{Vacuum stability with respect to $\Delta \kappa_{1,...13}$ terms}

\begin{figure*}
\sidecaption
\centering
\begin{minipage}[h]{0.33\linewidth}
\centering{\includegraphics[width=1\linewidth]{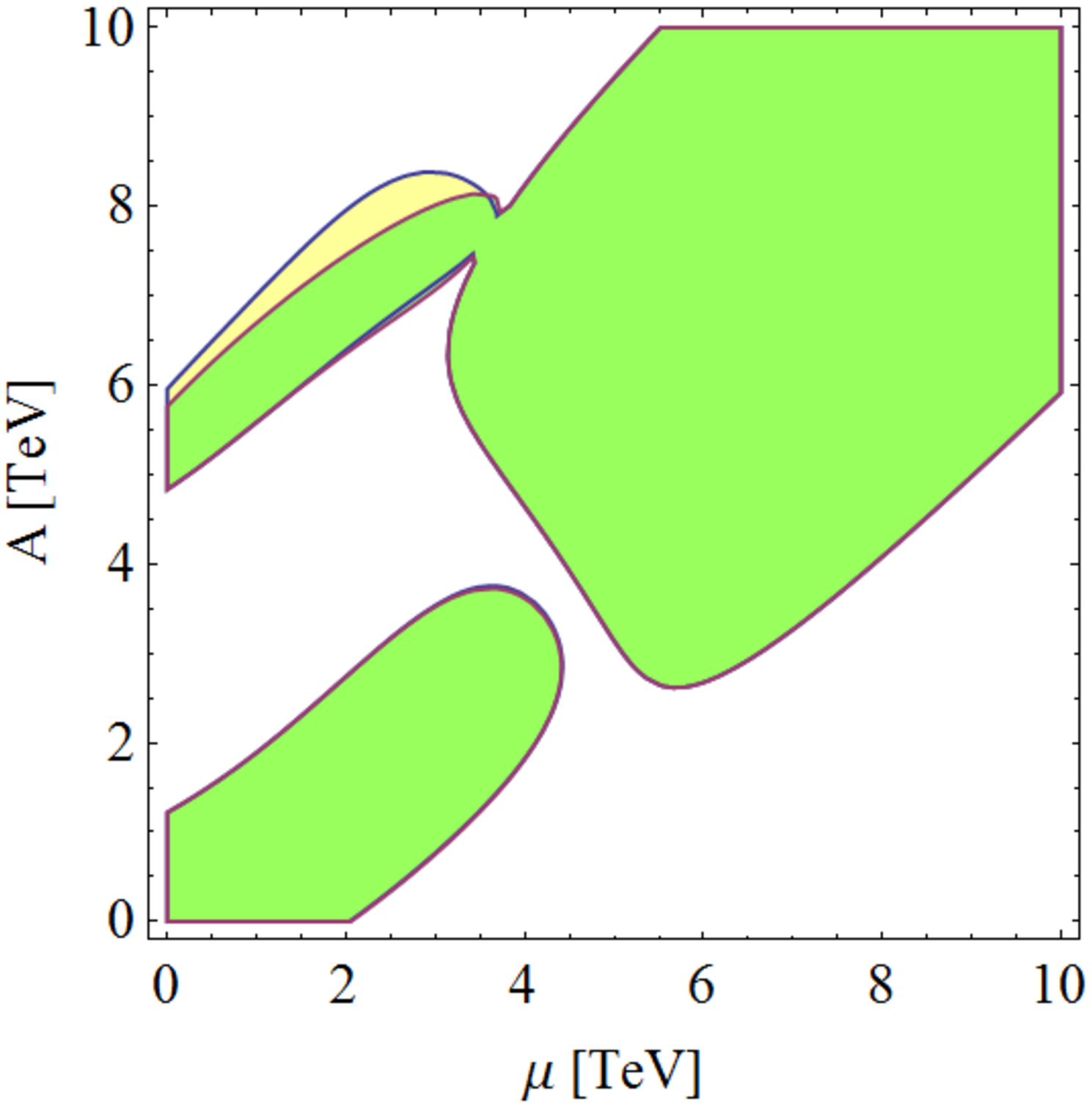}\\(a)}
\end{minipage}
\begin{minipage}[h]{0.33\linewidth}
\centering{\includegraphics[width=1\linewidth]{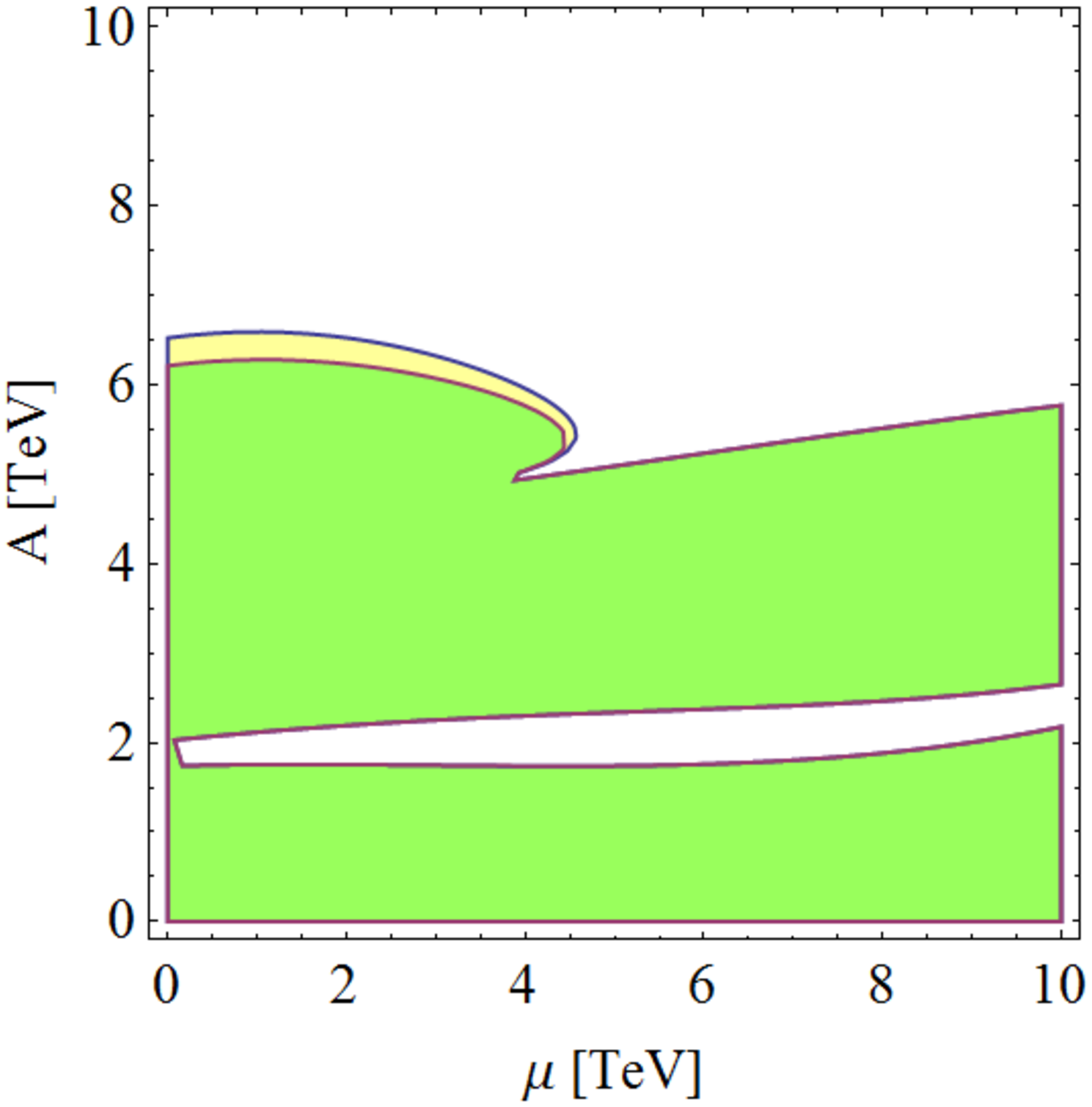}\\(b)}
\end{minipage}
\begin{minipage}[h]{0.33\linewidth}
\centering{\includegraphics[width=1\linewidth]{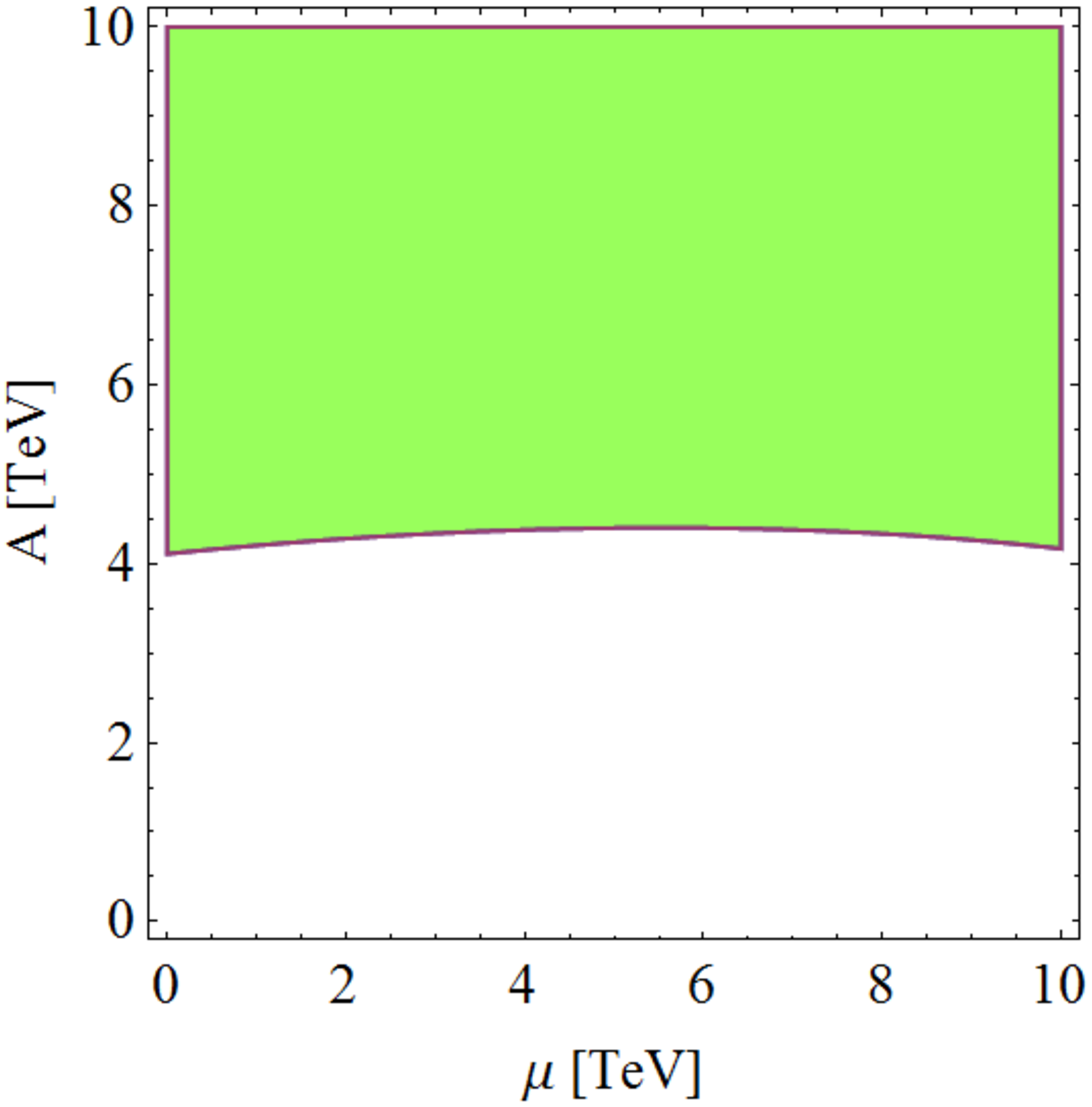}\\(c)}
\end{minipage}
\vspace*{2cm}       
\caption{The allowed regions defined by Eq. (\ref{det_cond}) for $m_h=$125 GeV and (a) -- $\tan \beta$=1, $M_S$=1.5 TeV, (b) -- $\tan \beta$=10, $M_S$=1.5 TeV and (c) $\tan \beta$=10, $M_S$=5 TeV.  Green area corresponds to an allowed parameter range for the Higgs potential $U=U^{(2)}+U^{(4)}+U^{(6)}$; yellow area corresponds to an allowed parameter range for the potential $U^{(2)}+U^{(4)}$ in the one-loop expansion of the effective Higgs potential, Eq.(\ref{U}).
Green and yellow areas mostly overlap.}
\label{fig-min}       
\end{figure*}

Precise measurements of the Higgs mass and the top quark mass are critical to determine the vacuum structure of the SM \cite{sm_vacuum}
which demonstrates a metastable vacuum. In the SM the scalar sector is single-field in the unitary gauge. In the SM extensions the situation is more complicated because the Higgs sector includes several fields.
Raises concerns the possibility of the existence of regions of parameter space where radiative corrections from dimension-six operators could spoil the vacuum stability. In the following we shall analyse the criteria when the local minimum $v=$246 GeV of the Higgs potential exists if the following
conditions 
are respected

\begin{eqnarray}
\label{extr_cond}
\left. \frac{\partial U}{\partial \phi_1^0} \right|_{\phi_i^0=v_i}=0, 
\qquad
\left. \frac{\partial U}{\partial \phi_2^0}\right|_{\phi_i^0=v_i}=0,\qquad
\label{det_cond}
\Delta=
\left| \left|
\begin{array}{cc}
U^{\prime \prime}_{\phi_1^0 \phi_1^0} &  U^{\prime \prime}_{\phi_1^0 \phi_2^0}\\
 U^{\prime \prime}_{\phi_2^0 \phi_1^0} & U^{\prime \prime}_{\phi_2^0 \phi_2^0}
\end{array}
\right| \right|_{\phi_i^0=v_i} >0, \qquad
\label{min_cond}
U^{\prime \prime}_{\phi_1^0 \phi_1^0} \Big|_{\phi_i^0=v_i}>0.
\end{eqnarray}
Rather compact form for $\Delta$ when $\kappa_i$=0 can be written as 
\begin{eqnarray}
\label{det}
\Delta &=&
	-[-2 {\rm Re}\mu_{12}^2 + v^2 (3 \lambda_6 c_\beta^2 + 3 \lambda_7 c_\beta^2 + \lambda_{345} s_{2 \beta})]^2 \\
	&-&
	[v^2 (-\lambda_7 + 3 \lambda_6 \cot^2 \beta + 4 \lambda_1 \cot^3 \beta) + 2 {\rm Re}\mu_{12}^2 \csc^2 \beta] \nonumber \\
	&\times&
	[v^2 (-4 \lambda_2 - 3 \lambda_7 \cot \beta + \lambda_6 \cot^3 \beta) - 
   2 {\rm Re}\mu_{12}^2 \cot \beta \csc^2 \beta] s_\beta^4 \tan \beta >0, \nonumber
\end{eqnarray}
where ${\rm Re}\mu_{12}^2$ is defined by Eq. (\ref{re12}). Allowed domains on the plane 
$(\mu,A)$ for fixed values of $M_S$ and $\tan \beta$ are shown in Fig. \ref{fig-min}.
They are reconstructed for the case when the lightest CP-even scalar mass $m_h=$125 GeV and masses of 
other Higgs bosons are positively defined.
The green region (it is superimposed on the yellow region of approximately the same shape) includes an allowed $A$ and $\mu$ range if the Higgs potential terms $U^{(2)}+U^{(4)}$ (see Eq.(\ref{U})) are extended by $U^{(6)}$ operators, Eq. (\ref{U6}), and the yellow area represents the allowed region for $U^{(2)}+U^{(4)}$ terms only (green and yellow domains rather precisely overlap excluding yellow stripes). Visible deviations of the contours for the dimension-four potential are observed if the values of $A_t$ respect the conditions of Eq. (\ref{6-dim-cond}).

\begin{figure*}
\centering
\begin{minipage}[h]{0.45\linewidth}
\centering{\includegraphics[width=0.9\linewidth]{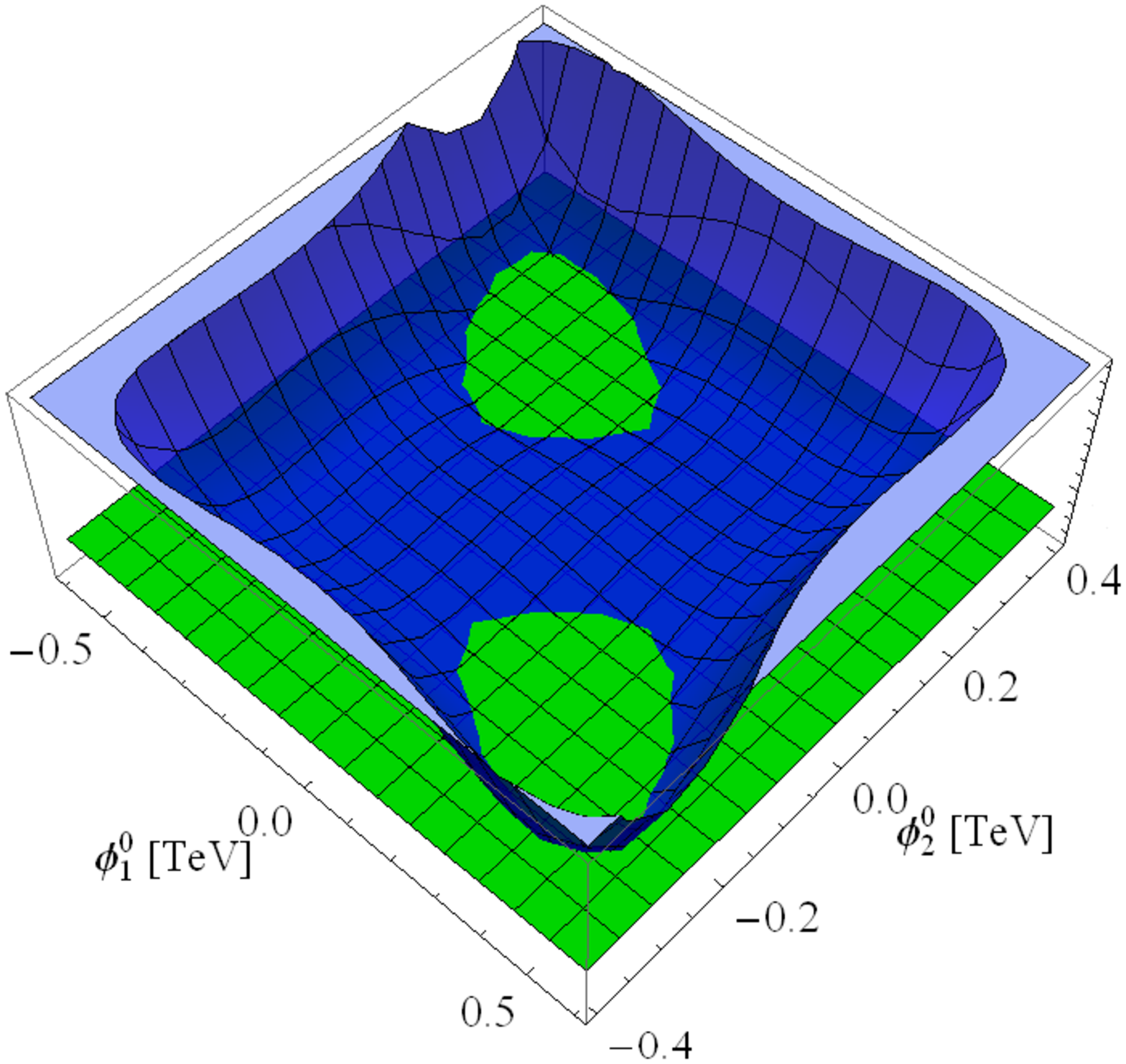}\\(a)}
\end{minipage}
\begin{minipage}[h]{0.45\linewidth}
\centering{\includegraphics[width=0.9\linewidth]{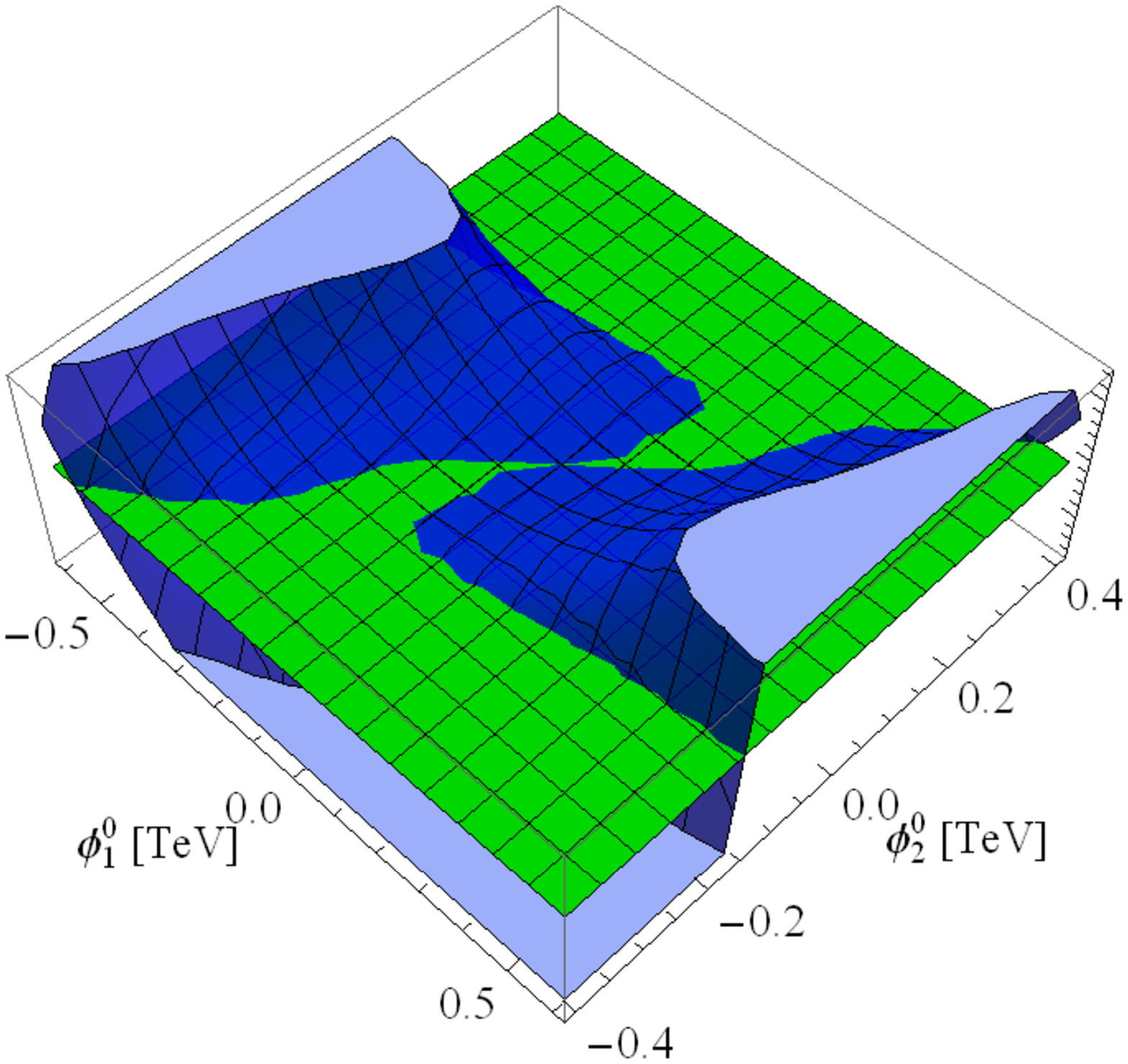}\\(b)}
\end{minipage}
\caption{
Higgs potential forms, see (\ref{U}), for $M_S$=1.5 TeV, $\tan \beta$=1 and
(a) -- $A=\mu$=1.5 TeV,
(b) -- $A$=7.5 TeV, $\mu$=2 TeV, see also Fig. \ref{fig-min}. The plane $U=$0 is shown in green colour.
}
\label{fig-Uforms}       
\end{figure*}

Configuration for $U^{(2)}+U^{(4)}+U^{(6)}$ in the case of "good"\vspace{0mm} ($A$, $\mu$) parameters for the effective potential decomposition is shown in Fig. \ref{fig-Uforms}(a). The potential has two minima and is positively defined at a large $\phi_i^0$. So in this case the vacuum is true and stable. But if $A$ and $\mu$ are chosen from the yellow stripes of Fig. \ref{fig-min} (a), the conditions defined by Eq. (\ref{extr_cond}) are not respected for the potential $U^{(2)}+U^{(4)}+U^{(6)}$, two minima degenerate to a gully which forms a saddle configuration at the origin, see Fig. \ref{fig-Uforms}(b). In the case 
$\Delta=0$, a gully forms the flat direction when the stationary point at the origin is degenerate and nonisolated \cite{tmf}.

In the considered benchmark scenarios $M_S<1.5$ TeV, $\mu<M_S$ (except \textit{$\tau$-phobic} scenario) and $A_t=X_t+\mu/\tan \beta$, where $X_t \sim {\cal O}(1) M_S$, so large values of $A_t$ when the difference between various radiative corrections is appreciable can be achieved only for extremely small $\tan \beta$ region, where additional contributions are expected.
The minimum of the potential disappears only at low $\tan \beta$ for \textit{light stop} and $\tau$-{\it phobic} scenarios while other demonstrate remarkable stability with respect to radiative corrections.

\section{Summary}
In the above examples it is evident that dimension-six effective operators moderately contribute as a rule, but in some scenarios they could mangle rather strongly the observables of the Higgs system with the dimension-four potential terms in the fields. Regions of the parameter space consistent with $m_h=$125 GeV change significantly, especially at low $\tan \beta$. At the same time in the regime of large MSSM parameters $A_t$,$\mu$ of the order of several TeV or more, moderate tuning is needed to combine consistently the low mass CP-odd state with the CP-even state at 125 GeV and heavy enough charged Higgs boson which is accompanied by the second CP-even state in the same mass range. The alignment limit which is required to match consistently all the decay channels of 125 GeV scalar is easily reached when the 125 GeV state is identified as the lightest CP-even boson $h$ and is hardly possible if it is identified as the heavier mass state $H$. Regimes with tachyonic states unsuitable for identification may appear.

Untrivial question of pertirbative unitarity conditions in the large $A_t$,$\mu$ regime deserves a separate study. In the standard procedure of analysis tree-level scattering matrix for all physical states is constructed in the mass basis and diagonalized imposing then constraints on the eigenvalues \cite{unitarity_1}. This procedure can be performed \cite{unitarity_2} in the two-Higgs doublet model basis of $SU(2)$ states, Eq.(\ref{dublets}), if it is related to the mass basis by a unitary transformation. Using specific non-linear symbolic formulae which express $\lambda_i$ through masses of bosons and mixing angles in the scalar sector \cite{specific} the parameter space can be directly constrained by imposing some meaningful limitations on the eigenvalues of the MSSM scattering matrix.     

{\it Acknowledgements} \hskip 2mm The authors are grateful to E. Boos, L.Dudko, A. Nikitenko and I. Volobuev for useful comments. This work was supported by Russian Science Foundation Grant No. 16-12-10280.

\end{document}